# AN OVERVIEW OF THE SECURITY CONCERNS IN ENTERPRISE CLOUD COMPUTING

Anthony Bisong[1] and Syed (Shawon) M. Rahman[2]

[1]Ph.D. Student, Capella University
225 South 6th Street, 9th Floor Minneapolis, MN 55402, USA
Email: abisong@gmail.com

[2]Assistant Professor of Computer Science, University of Hawaii-Hilo, Hilo, HI, USA
and Adjunct Faculty, Capella University, Minneapolis, MN 55402,USA
Email: SRahman@Hawaii.edu

## Abstract

*Deploying cloud computing in an enterprise infrastructure bring significant security concerns. Successful implementation of cloud computing in an enterprise requires proper planning and understanding of emerging risks, threats, vulnerabilities, and possible countermeasures. We believe enterprise should analyze the company/organization security risks, threats, and available countermeasures before adopting this technology. In this paper, we have discussed security risks and concerns in cloud computing and enlightened steps that an enterprise can take to reduce security risks and protect their resources. We have also explained cloud computing strengths/benefits, weaknesses, and applicable areas in information risk management.*

## 1.0 Introduction

This paper discusses the cloud computing security concerns and the security risk associated with enterprise cloud computing including its threats, risk and vulnerability. Throughout the years, organizations have experienced and will continue to experience in this cloud computing era numerous system losses which will have a direct impact on their most valuable asset, information (Otero, Otero, Qureshi, 2010) and its protection is utmost important to all organizations. There have been publicized attacks on cloud computing providers and this paper discusses recommended steps to handle cloud security, issues to clarify before adopting cloud computing, the need for a governance strategy and good governance technology, cloud computing strengths, weaknesses, analyzes the benefits and costs of cloud computing in information security management.

Cloud computing is continuously evolving and there are several major cloud computing providers such as Amazon, Google, Microsoft, Yahoo and several others who are providing services such as Software-as-a-Service (SaaS), Platform-as-a-Service (PaaS), Storage-as-a-Service and Infastructure-as-a-Service (IaaS) and this paper has discussed some of the services being provided. There are many scholarly researches, articles and periodicals on cloud computing security concern out there. Security researchers and professionals are working on security risks, potential threats, vulnerabilities, and possible countermeasure in enterprise cloud computing constantly.

## 2.0 Background Study

Enterprises are starting to look into cloud computing technology as a way to cut down on cost and increase profitability, because across all industries "CIOs are under continuous pressure to reduce capital assets, headcounts, and support costs, and cloud systems give them a way to meet those goals" (Brendl, 2010).

There are many definitions of cloud computing and the most comprehensive definition available is by Brendl (2010) who defined cloud computing as "collections of IT resources (servers, databases, and applications) which are available on an on-demand basis, provided by a service company, available through the internet, and provide resource pooling among multiple users." Figure 1. shows what is available to enterprises in the cloud.

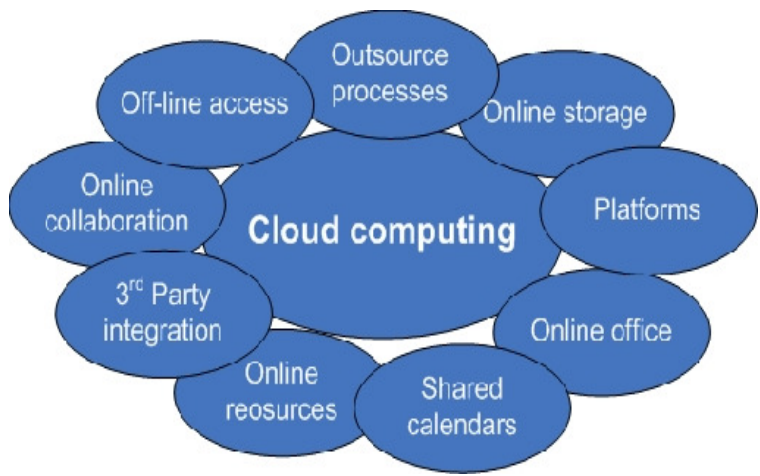

Figure 1. Cloud Computing Resources (CloudTweaks, 2010)

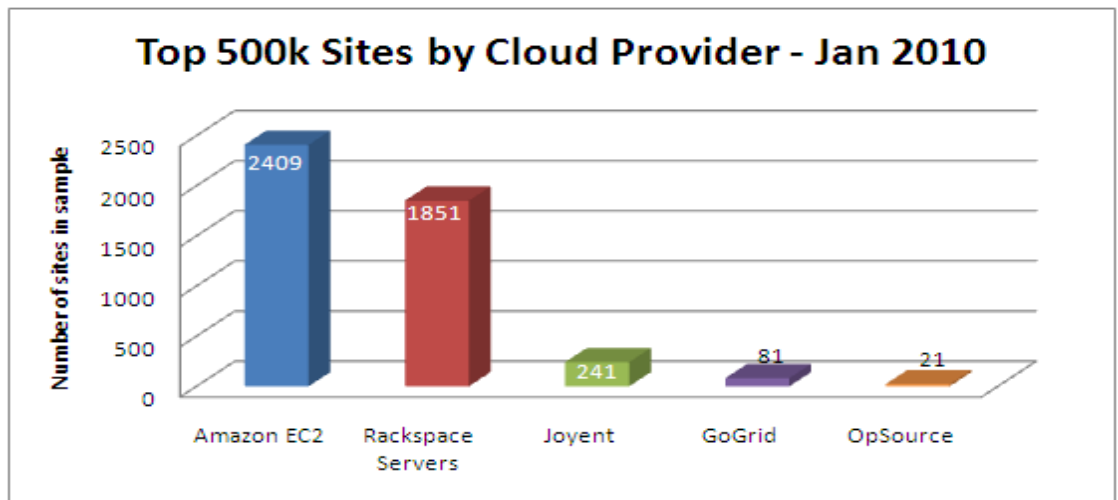

Figure 2. Cloud Providers. Top 500k sites by Cloud Provider - April 2010 (CloudTweaks, 2010).

Brendl (2010) went on to say that due to the potential profitability of cloud computing to save enterprises money and increase the bottom line "CIOs are looking for any and all opportunities to move internal company systems to external cloud systems because cloud systems reduce capital assets, IT maintenance costs, and direct labor costs".  Figure 2. shows the top 500k sites by the major cloud providers.

## 2.1 Cloud Computing Growth

Cloud computing is a combination of several key technologies that have evolved and matured over the years (see Figure 3.).  This evolution to present day cloud computing includes a combination of open API's, storage, computing, infrastructure as shown in Figure 3.

The "cost associativity" formulae as shown in Formula 1. (Armbrust, Fox, Griffith, Joseph, Katz, Konwinski and et al., 2009) can be used to compute the profitability of cloud computing.  For example using 1000 Amazon EC2 machines for 1 hour costs the same as using 1 traditional non cloud machine for 1000 hours (Armbrust et al., 2009).

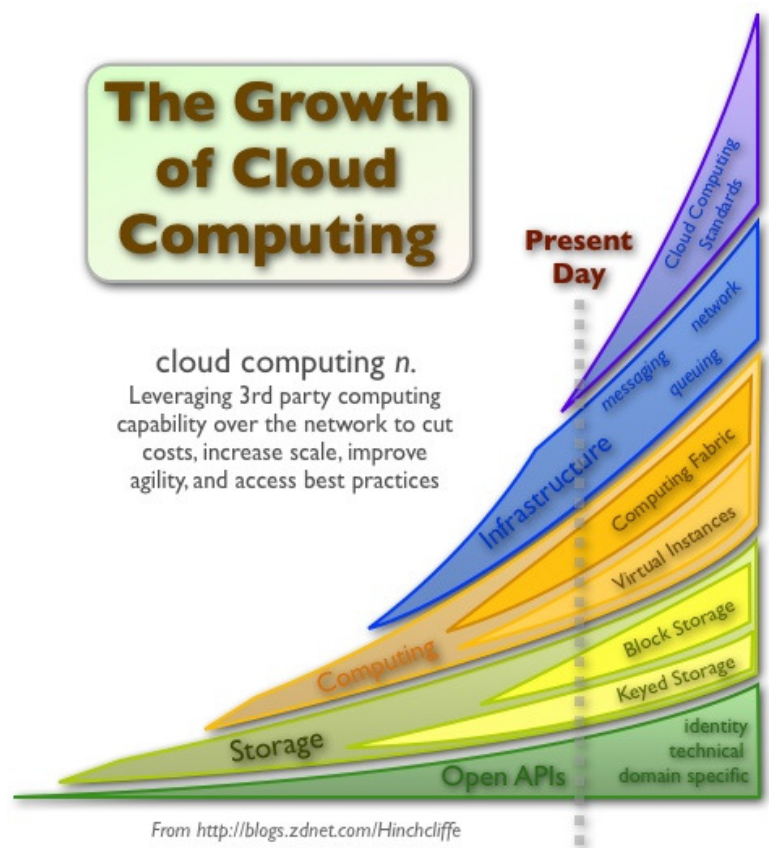

The Profitability of cloud computing can be explained in the "cost associativity" formulae shown in Formula 1., the left-hand side multiplies the net revenue per user-hour by the number of user-hours, giving the expected profit from using cloud computing while the right-hand side performs the same calculation for a fixed-capacity datacenter by factoring in the average utilization, including nonpeak workloads, of the datacenter; whichever side is greater represents the opportunity for higher profit" (Armbrust et al., 2009).  Ambrust et al. (2009, p. 10-11) gave example on elasticity with calculations on the potentials of cloud computing savings and cost reduction:

Figure 3.  The growth of Cloud Computing. (Hinchcliffe, 2009)

$$\text{UserHours}_{cloud} \times (\text{revenue} - \text{Cost}_{cloud}) \geq \text{UserHours}_{datacenter} \times (\text{revenue} - \frac{\text{Cost}_{datacenter}}{\text{Utilization}})$$

Formula 1. (Armbrust et al., 2009 p. 2).

## 2.2 Cloud Computing Example

There are several major cloud computing providers including Amazon, Google, Salesforce, Yahoo, Microsoft and others that are providing cloud computing services (Figure 4. shows current cloud providers).

Cloud computing providers provide a variety of services to the customers and these services include e-mails, storage, software-as-a-services, infrastructure-as-a-services etc.

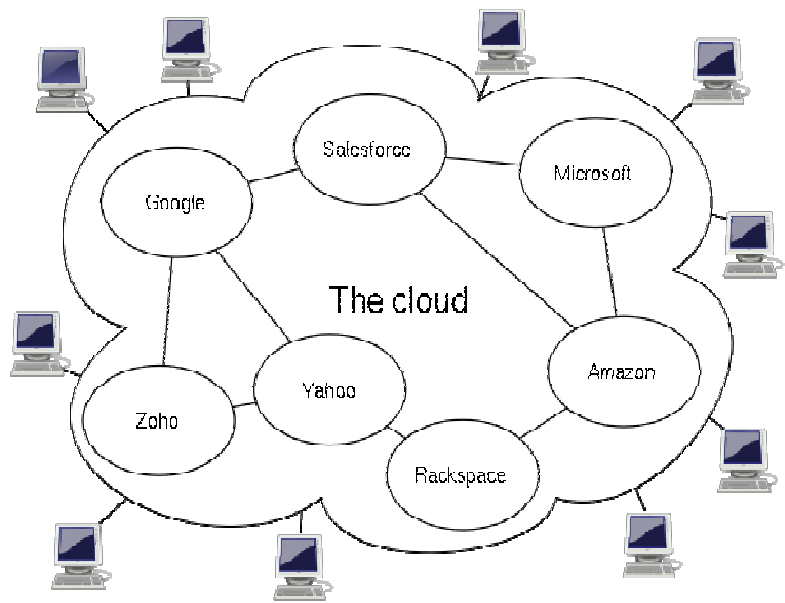

The attractiveness of cloud computing is not only to large enterprises but also entrepreneurs, startups, medium companies and small companies would benefit greatly and they will have a new alternative and opportunities that is not available to them in the past that would save them millions of dollars because with cloud computing they will have the choice to only rent the necessary computing power, storage space and communication capacity from a large cloud computing provider that has all of these assets connected to the Internet (Smith, 2009).

Figure 4. Cloud Computing Overview (CloudTweaks, 2010).

Companies "can pay only for the volume of these services that they use, they can quickly add or subtract resources from their order, and they never have to take possession of the hardware and all of the technical support headaches associated with it" (Smith, 2009). Table 1 shows the pay per use competitive matrix of some of the major cloud computing providers for infrastructure as a service (IaaS), platform as a service (PaaS).

Smith (2009) gave an example of the tremendous benefits of cloud computing to a company, how a startup company call Animoto that allow people to turn a series of photographs into a simple movie with a nice sound track in the background and have them online to share with friends and family become a poster child for the cloud computing concept; when the online photo to movie application became available on the internet, suddenly over a three-day period Animoto registration increased from 25,000 to 250,000 users and as a result they ramped up their usage of Amazon cloud computers from 24 machines to nearly 5,000 machines within a week. This capability before cloud computing would have been near impossible and it could have cost the startup company millions of dollars and several months of time and effort to achieve without cloud computing.

Table 1. Competitive matrix. (CloudTweaks, 2010) diagrams

| Provider | IaaS | PaaS | Compute Billing Model | Storage Billing Model | Relational Database Service | Hybrid capabilities |
|---|---|---|---|---|---|---|
| **Windows Azure** | No | Yes (.Note, Java, Ruby, Python, PHP) | Pay-per-use | Pay-per-use | Yes (SQL Server) | Yes (on-premise to cloud) |
| **Amazon Web Services** | Yes | No | Pay-per-use | Pay-per-use | Yes (MySQL-based) | Yes (via third-party tools) |
| **Rackspace** | Yes | Yes (LAMP, .Net (PaaS) | Pay-per-use (IaaS); Monthly (PaaS) | Pay-per-use (IaaS); Included in monthly fee (PaaS) | Yes (FathomDB) | No (but dedicated resources to cloud planned) |
| **Joyent** | Yes | Yes (Java, Ruby, Python, PHP) | Monthly (IaaS); PaaS pricing not announced | Included in monthly fee (IaaS) | No | Yes (on-premise to cloud) |
| **Google** | No | Yes (Python, Java) | Pay-per-use | Pay-per-use | No | No |
| **GoGrid** | Yes | No | Pay-per-use or pre-paid | Included with each instance | No | Yes (dedicated resources to cloud) |

With all these cloud computing capabilities and potential to save cost, large enterprises and others should step back, move cautiously and analyze the security risks and concerns associated with cloud computing before adapting the technology.

## 2.3 Cost Minimization

Cost minimization of cloud computing to enterprises can be explained in cloud computing elasticity capability. Assume our service has a predictable daily demand where the peak requires 500 servers at noon but the trough requires only 100 servers at midnight, as shown in Figure 2(a). As long as the average utilization over a whole day is 300 servers, the actual utilization over the whole day (shaded area under the curve) is 300 x 24 = 7200 server-hours; but since we must provision to the peak of 500 servers, we pay for 500 x 24 = 12000 server-hours, a factor of 1.7 more than what is needed. Therefore, as long as the pay-as-you-go cost per server-hour over 3 years is less than 1.7 times the cost of buying the server, we can save money using utility computing (Armbrust et al., 2009 p. 10).

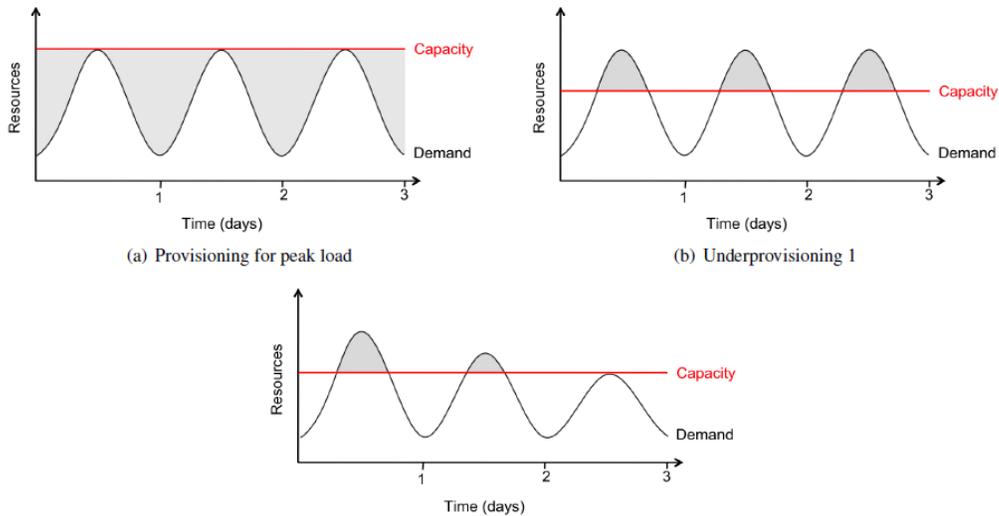

Figure 5. Provisioning for peak load and underprovisiong (Armbrust et al., 2009 p. 11).

In Figure 5: (a) Even if peak load can be correctly anticipated, without elasticity we waste resources (shaded area) during nonpeak times. (b) Underprovisioning case 1: potential revenue from users not served (shaded area) is sacrificed. (c) Underprovisioning case 2: some users desert the site permanently after experiencing poor service; this attrition and possible negative press result in a permanent loss of a portion of the revenue stream (Armbrust et al., 2009).

### 3.0 Security Threats, Risks, and Vulnerabilities

With the increasing popularity of enterprise cloud computing and its public connectivity via the internet it is the next frontier for viruses, worms, hackers and cyber-terrorists to start probing and attacking. Many enterprises are seriously looking into cloud computing to save cost, in the not too distance future cloud computing adoption rate will skyrocket and cloud computing vulnerability to viruses, worms, hackers and cyber attacks will increase because organized criminals, terrorist and hostile nations would see this as a new frontier to try to steal private information, disrupt services and course harm to the enterprise cloud computing network. Cloud computing security risk incident has happened when Google a major cloud computing and Software as a Service (SaaS) provider had its systems attacked and hacked; the cyber-forensics has been traced to the attacks coming from China (Markoff, Barboza, 2010).

With cloud computing, physical location of data are spread across geographic area that could span over continents, countries or regions. One of the top security concerns of enterprises are the physical location of the data that are being stored in the cloud especially if they are located in another country because the laws of the host country of the equipment apply to the data on the machines (Smith, 2009) and that could be a big issue if the host country does not have adequate laws to protect sensitive data or if the host nation becomes hostile or when the government of the hosting nation changes and become unfriendly.

There have been instances where there has been a complete blackout of entire cloud services and making it unavailable for hours and even days due to bugs (Smith, 2009). Google's Gmail went down for two hours, Ctrix's GoToMeeting and GoToWebinar were temporarily unavailable, Amazon.com's Simple Storage Service was "out of commission for excruciating eight hours" (Hoover, 2008). Imagine an enterprise that completely depends on a cloud computing service provider whose system had been disrupted for hours or days, the lost of business could be catastrophic.

## 3.1 Threats

Cloud computing faces just as much security threats that are currently found in the existing computing platforms, networks, intranets, internets in enterprises. These threats, risk vulnerabilities come in various forms. The Cloud Security Alliance (Cloud Computing Alliance, 2010) did a research on the threats facing cloud computing and it identified the flowing seven major threats:

- ♦ Abuse and Nefarious Use of Cloud Computing
- ♦ Insecure Application Programming Interfaces
- ♦ Malicious Insiders
- ♦ Shared Technology Vulnerabilities
- ♦ Data Loss/Leakage
- ♦ Account, Service & Traffic Hijacking
- ♦ Unknown Risk Profile

## 3.2 Risks

Risk according to SAN Institute "is the potential harm that may arise from some current process or from some future event." In IT security, risk management is the process in which we understand and respond to factors that may lead to a failure in the confidentiality, integrity or availability of an information system (SAN Institute); the IT security risk is the harm to a process or the related information resulting from some purposeful or accidental event that negatively impacts the process or the related information (SANS Institute).

Moving to the cloud presents the enterprise with a number of risks and that include securing critical information like the protection of intellectual property, trade secrets, personally identifiable information that could fall into the wrong hands. Making sensitive information available on the internet requires a considerable investment in security controls and monitoring of access to the contents. In the cloud environment, the enterprise may have little or no visibility to storage and backup processes and little or no physical access to storage devices by the cloud computing provider. And, because the data from multiple customers may be stored in a single repository, forensic inspection of the storage media and a proper understanding of file access and deletion will be a significant challenge (Information Security Magazine, 2009).

## 3.3 Vulnerability

According to Pfleenger (2006) vulnerability "is a weakness in the security system" that could be exploited to cause harm. Enterprise cloud computing is just as vulnerable as any other technology that uses the public internet for connectivity. The vulnerability includes eavesdropping, hacking, cracking, malicious attacks and outages. Moving your data to a cloud service is just like "putting all your eggs in one basket" (Perez, 2009) and in early 2009 social bookmarking site Ma.gnolia experienced a server crash in which it lost massive data of its users that its bookmarking services was shut down permanently.

Research has shown that it is possible for attackers to precisely map where a target's data is physically located within the "cloud" and use various tricks to gather intelligence (Talbot, 2009, p. 1). Another vulnerability to an attack is the use of denial-of-service attack and it has been found out that if an attacker is on the same cloud servers as his victim, a conventional denial-of-service attack can be initiated by amping up his resource usage all at once (Talbot, 2009, p. 5).

Researchers at the University of California at San Diego and at M.I.T. say they can buy cloud services from Amazon and place a virtual machine on the same physical machine as a target application and once there, they can use their virtual machine's access to the shared resources of the physical machine to steal data such as passwords (Greene, 2009). This technique the researchers said is experimental and doesn't work all the time, but it indicates that service providers' clouds are susceptible to new types of attacks not seen before. And while they attacked was carried out inside Amazon's EC2 cloud, they say their method would work equally well with other cloud providers. (Greene, 2009).

The researchers went on to say that a way around the weakness they found in Amazon's EC2 is for customers to insist that their cloud machines are placed on physical machines that only they can access or that they and trusted third parties can access (Greene, 2009). This solution will likely be at a price premium because part of the economy of cloud services is maximizing use of physical servers by efficiently loading them up with cloud machines (Greene, 2009) and locating the cloud datacenter where the utility price is the cheapest.

The work by the researchers highlights that clouds and the virtual environments they employ are relatively new; as a result they still draw the attention of attackers bent on finding and exploiting unexplored vulnerabilities (Greene, 2009). This doesn't mean that cloud services are unsafe and shouldn't be used (Greene, 2009).

In defending cloud computing security Edwards (2010) said that by using cloud computing as a "thin client technology, businesses can limit exposure threats posed by data-crammed laptops and backups. There will be more efficient security software because with cloud computing software vendors will be driven to fix inefficient security approaches that burn up resources (Edwards, 2010). The cloud will be a better anti-virus detection and the University of Michigan researchers has found out that if anti-virus software tools were moved from a PC to the cloud they could detect 35 percent more recent viruses than a single anti-virus program on a personal computer (Edwards, 2010). The bottom line is that businesses should treat clouds with a certain amount of suspicion; they should assess the risk the cloud service represents and only commit data to such services that can tolerate that risk" (Greene, 2009).

## 4. Cloud Computation Implementation Guidelines

### 4.1 Steps to Cloud Security

Edwards (2009) stated that, with the security risk and vulnerability in the enterprise cloud computing that are being discovered enterprises that want to proceed with cloud computing should, use the following steps to verify and understand cloud security provided by a cloud provider:

- ♦ **Understand the cloud** by realizing how the cloud's uniquely loose structure affects the security of data sent into it. This can be done by having an in-depth understanding of how cloud computing transmit and handles data.
- ♦ **Demand Transparency** by making sure that the cloud provider can supply detailed information on its security architecture and is willing to accept regular security audit. The regular security audit should be from an independent body or federal agency.
- ♦ **Reinforce Internal Security** by making sure that the cloud provider's internal security technologies and practices including firewalls and user access controls are very strong and can mesh very well with the cloud security measures.
- ♦ **Consider the Legal Implications** by knowing how the laws and regulations will affect what you send into the cloud.
- ♦ **Pay attention** by constantly monitoring any development or changes in the cloud technologies and practices that may impact your data's security.

### 4.2 Issues to Clarify Before Adopting Cloud Computing

Gartner, Inc., the world's leading information technology research and advisory company, has identified seven security concerns that an enterprise cloud computing user should address with cloud computing providers (Edwards, 2009) before adopting:

- ♦ **User Access.** Ask providers for specific information on the hiring and oversight of privileged administrators and the controls over their access to information. Major companies should demand and enforce their own hiring criteria for personnel that will operate their cloud computing environments.
- ♦ **Regulatory Compliance.** Make sure your provider is willing to submit to external audits and security certifications.
- ♦ **Data location.** Enterprises should require that the cloud computing provider store and process data in specific jurisdictions and should obey the privacy rules of those jurisdictions.

- ♦ **Data Segregation.** Find out what is done to segregate your data, and ask for proof that encryption schemes are deployed and are effective.
- ♦ **Disaster Recovery Verification**. Know what will happen if disaster strikes by asking whether your provider will be able to completely restore your data and service, and find out how long it will take.
- ♦ **Disaster Recovery.** Ask the provider for a contractual commitment to support specific types of investigations, such as the research involved in the discovery phase of a lawsuit, and verify that the provider has successfully supported such activities in the past. Without evidence, don't assume that it can do so.
- ♦ **Long-term Viability.** Ask prospective providers how you would get your data back if they were to fail or be acquired, and find out if the data would be in a format that you could easily import into a replacement application.

### 4.3 Need for a Governance Strategy and Good Governance Technology

Moving into the cloud computing requires a good governance strategy and a good governance technology (Kobielus, 2009). Interest in governance has been revitalize because trust is being extended to a cloud provider across premise and across corporate boundaries (Kobielus, 2009, p. 26). A cloud computing governance function requires active management participation, the proper forum to make IT related decisions, and effective communication between the IT organization and the company's management team (Maches, 2010). Maches (2010) proposed cloud risk management be included in the cloud computing governance function that requires risk awareness by senior corporate officers, a clear understanding of the enterprise's appetite for risk, understanding of compliance requirements, transparency about the significant risks to the enterprise and embedding of risk management responsibilities into the IT organization.

## 5.0 Cloud Computing Strengths, weaknesses, and Application Areas in Information Risk Management

### 5.1 Cloud Computing Strengths/Benefits

The strength of cloud computing in information risk management is the ability to manage risk more effectively from a centralize point. Security updates and new patches can be applied more effectively thereby allowing business continuity in an event of a security hole.

### 5.2 Weaknesses

Cloud computing weakness include list of issues such as the security and privacy of business data being hosted in remote 3rd party data centers, being lock-in to a platform, reliability/performance concerns, and the fears of making the wrong decision before the industry begins to mature (Hinchcliffe, 2009).

## 5.3 The benefits and costs of Cloud Computing in information security management

According to Bendandi (2009, p. 7) the top security benefits of cloud computing includes:

♦   The security and benefits of scale that all kinds of security measures are cheaper when implemented on a large scale including all kinds of defensive measures such as filtering, patch management, hardening of virtual machine instances and hypersivors, etc.  The benefits of scale also include multiple locations, edge networks (content delivered or processed closer to its destination), timeliness of response to incidents and centralized threat management.

♦   Security as a market differentiator that give cloud providers a strong driver to improve security practices and many cloud customers will buy on the basis of the reputation for confidentiality, integrity and resilient of and the security services offered by a provider

♦   Large cloud providers will offer a standardized, opened interface to manage security thereby opening a market for security services.

♦   Rapid and smart scaling of resources where cloud provider dynamically reallocate resources for filtering, traffic shaping, authentication, encryption and defensive measures such as distributed denial-of-service (DDoS) attack

♦   Audit and evidence-gathering where dedicated pay-per-use forensic images of virtual machines are accessible without taking infrastructure offline and it provide cost-effective storage for logs allowing comprehensive logging without compromising performance.

The cost of cloud computing in information security management includes the costs of migrating, implementing, integrating, training, and redesigning.  Also it includes the cost of training supporting people in the new processes. The new architecture could generate new security holes and issues during redesigning and deploying the implementation thereby driving cost up.  In the application areas in information risk management, cloud computing is commercially viable alternative for enterprises in search of a cost-effective storage and server solution. (Waxer, 2010). Gartner Inc. predicts that by 2012, 80 percent of Fortune 1000 enterprises will pay for some cloud-computing service (Waxer, 2010), while 30 percent of them will pay for cloud-computing infrastructure. While the technology has its fair share of drawbacks (such as privacy and security concerns), an undeniable potential benefit is turning a lot skeptics into enthusiasts (Waxer, 2010).

## 6.0 Recommendations

The following recommendations and strategies put forward by Indiana University(2009) intended to assist its departments and units in their approach to evaluating the prudence and feasibility of leveraging cloud services can also be used in accessing cloud computing in enterprises.

♦ Risk/benefit analysis: Units considering university services that may be delivered using cloud technology, or new services provided by cloud technology, must indentify and understand the risks and benefits of the service. Recognize that vendor security failures will potentially involve or at least reflect on the university. Consider the security and privacy objectives of confidentiality, integrity, availability, use control, and availability, and determine what would happen if these objectives were not met. Honestly compare costs of the internal and external services, including costs to manage the vendor relationship, and costs of integrating the service with existing internal services and processes.

♦ Consultation: Consult with appropriate data stewards, process owners, stakeholders, and subject matter experts during the evaluation process. Also, consult with Purchasing, the General Counsel's Office, the University Information Policy Office, and the University Information Security Office.

♦ Lower risk candidates: When considering university services that may be delivered using cloud technology, ideal candidates will be those that are non-critical to operations, involve public information, and otherwise would require significant internal infrastructure or investment to deliver or continue delivering internally. These are likely to represent the best opportunities for maximizing benefit while minimizing risk.

♦ Higher risk candidates: University services that are critical to the operation of the university or involve differentiating or core competencies, and/or involve restricted, or critical information or intellectual property, are necessarily higher risk candidates and require careful scrutiny.

♦ Consider "internal cloud" alternatives: Due to the decentralized nature of the university, some duplication of effort is inevitable. Units should consider leveraging internal cloud-like services when looking for ways to reduce cost, e.g., units managing their own email servers and/or server hardware should consider migrating to the institutional email solutions and/or a virtual server solution (i.e., Intelligent Infrastructure). "Large enterprises should generally avoid placing sensitive information in public clouds, but concentrate on building internal cloud and hybrid cloud capabilities in the near term," (Dan Blum, "Cloud Computing Security in the Enterprise," Burton Group, July 15, 2009).

♦ Vendor agreement: In all cases, strive to obtain a contract or service level agreement with the vendor. For non-critical services involving public data, it may be possible to leverage a cloud service without such an agreement if the vendor is willing to provide adequate assurances; however, services critical to the university and/or those involving more sensitive data (i.e., restricted or critical) must not be provided by a cloud vendor without an appropriate agreement in place. Purchasing, the General Counsel's Office, the University Information Policy Office, and the University Information Security Office must be consulted when drafting such agreements.

♦ Proportionality of safeguards: Vendor physical, technical, and administrative safeguards should be equal to or better than those in place internally for similar services and information. Areas to explore with the vendor include privileged user access, regulatory compliance, data location, data segregation, recovery/data availability, change management, user provisioning and de-provisioning, personnel practices, incident response plans, and investigative/management support, as well as the issues identified in the previous section. Scrutinize any gaps identified.

♦ Due diligence: Due diligence should be conducted to determine the viability of the vendor/service provider. Consider such factors as vendor reputation, transparency, references, financial (means and resources), and independent third-party assessments of vendor safeguards and processes.

♦ Exit strategy: Cloud services should not be engaged without developing an exit strategy for disengaging from the vendor or service and integrating the service into business continuity and disaster recovery plans. Be sure to determine how you would recover your data from the vendor, especially in cases where the vendor shuts down.

♦ Proportionality of analysis/evaluation: The depth of the above analysis and evaluation and the scope of risk mitigation measures and required vendor assurances must be proportional to the risk involved, as determined by the sensitivity level of the information involved and the criticality or value to the University of the Service involved.

## 7.0 Conclusion

Cloud computing is a combination of several key technologies that have evolved and matured over the years**.** Cloud computing has a potential for cost savings to the enterprises but the security risk are also enormous. Enterprise looking into cloud computing technology as a way to cut down on cost and increase profitability should seriously analyze the security risk of cloud computing.

The strength of cloud computing in information risk management is the ability to manage risk more effectively from a centralize point. Security updates and new patches can be applied more effectively thereby allowing business continuity in an event of a security hole. Cloud computing weakness include list of issues such as the security and privacy of business data being hosted in remote 3rd party data centers, being lock-in to a platform, reliability/performance concerns, and the fears of making the wrong decision before the industry begins to mature.

Enterprise should verify and understand cloud security, carefully analyze the security issues involved and plan for ways to resolve it before implementing the technology. Pilot projects should be setup and good governance should be put in place to effectively deal with security issues and concerns. We believe the move into the cloud computing should be planned and it should be gradual over a period of time.

## Authors

### Anthony Bisong

Anthony Eban Bisong is a Ph.D. Student at Capella University majoring in Information Assurance and Security . Anthony also works full time consulting as Senior Software Engineer and has worked on Information Technology projects at fortune 500 corporations using leading internet technologies. His experience and interest in information technology are vast and includes Cloud Computing; Smartphone and mobile technologies – Android, IPhone, JavaFX, MeeGo, WebOS; Information Security and Internet technology using Web 2.0 technologies.

### Syed (Shawon) M. Rahman

Syed (Shawon) Rahman is an Assistant Professor in the Department of Computer Science & Engineering at the University of Hawaii-Hilo and an adjunct faculty of Information Technology, Information Assurance & Security at the Capella University. Dr. Rahman's research interests include Software Engineering Education, Data Visualization, Data Modelling, Information Assurance & Security, Web Accessibility, and Software Testing & Quality Assurance. He has published more than 50 peer-reviewed papers. He is an active member of many professional organizations including ACM, ASEE, ASQ, IEEE, and UPE.